\def\year{2019}\relax
\documentclass[letterpaper]{article} %DO NOT CHANGE THIS
\usepackage{aaai19}  %Required
\usepackage{times}  %Required
\usepackage{helvet}  %Required
\usepackage{courier}  %Required
\usepackage{url}  %Required
\usepackage{graphicx}  %Required
\usepackage{booktabs,threeparttable,multirow} % For formal tables
\usepackage{subfigure,epsfig,bm,amsmath,amsfonts}
\begin{document}
\title{CIA: Towards a Unified Marketing Optimization Framework for e-Commerce Sponsored Search}
\author{
Hao Liu$^\dagger$, Qinyu Cao$^\ddagger$, Xinru Liao$^\ddagger$, Guang Qiu$^\ddagger$, Sheng Li$^\ddagger$, Jiming Chen$^\dagger$ \\
$^\dagger$Zhejiang University, Hangzhou 310027, China \\
$^\ddagger$Alibaba Group, 969 West Wenyi Road, Hangzhou, China \\
\{vectorliu,cjm\}@zju.edu.cn, \{qingyu.cqy, xinru.lxr, guang.qiug, junqian.ls\}@alibaba-inc.com	
}
% \author{% AAAI Press\\
% % Association for the Advancement of Artificial Intelligence\\
% % 2275 East Bayshore Road, Suite 160\\
% % Palo Alto, California 94303\\
% }
% Remove the copyright information in the footnote
\nocopyright
\maketitle
\begin{abstract}
As the largest e-commerce platform, Taobao helps advertisers reach billions of search queries each day via sponsored search, which has also contributed considerable revenue to the platform. An efficient bidding strategy to cater to diverse advertiser demands while balancing platform revenue and consumer experience is significant to a healthy and sustainable marketing ecosystem. In this paper we propose \emph{Customer Intelligent Agent (CIA)}, a bidding optimization framework which implements an impression-level bidding to reflect advertisers' conversion willingness and budget control. In this way, CIA is capable of fulfilling various e-commerce advertiser demands on different levels, such as Gross Merchandise Volume optimization, style comparison etc. Additionally, a replay based simulation system is designed to predict the performance of different take-rate. CIA unifies the benefits of three parties in the marketing ecosystem without changing the Generalized Second Price mechanism. Our extensive offline simulations and large-scale online experiments on \emph{Taobao Search Advertising (TSA)} platform verify the high effectiveness of the CIA framework. Moreover, CIA has been deployed online as a major bidding tool in TSA.
\end{abstract}

\section{Introduction}\label{sec:introduction}
% Sponsored search in Taobao
Sponsored search has provided considerable revenue for universal search engines such as Google, Bing and vertical e-commerce websites like Taobao and Amazon. The 2017 Tmall Double Eleven shopping festival sees total sales of $25.4$ billion \cite{Double11}, showing the prosperity of online shopping and associated advertising market. As the biggest e-commerce platform in China \cite{Edquid17}, more than $300$ million consumers in Taobao reach commodities via keyword search each day, bringing daily $10$ billion search queries and subsequent page views (PVs). Such massive PVs provide advertisers (mainly Taobao sellers) sufficient opportunities to expose their commodities and services. 

\begin{figure}
\centering
\includegraphics[width=3.0in]{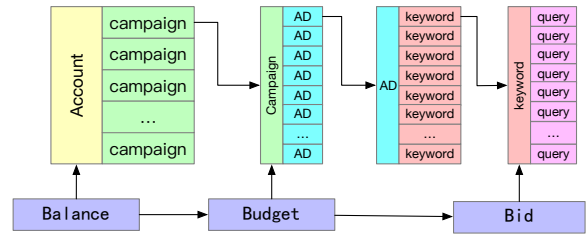}
\caption{Advertiser TSA account architecture.}\label{fig:account_architecture}
\end{figure}

\emph{Taobao Search Advertising (TSA)} is a suite of commercial advertising service \footnote{\url{http://zhitongche.taobao.com}}. Each advertiser maintains a tree-like TSA account as shown in Fig. \ref{fig:account_architecture}. The account has a balance and consists of several budget-limited marketing campaigns, while a campaign accommodates a set of ADs \footnote{In TSA, an advertising commodity or service is called an AD.} to be promoted. At the most fine-grained level, each AD specifies multiple relevant keywords and sets a fixed bid on each of them to compete for impressions.
Search queries initiated by consumers retrieve relevant ADs through exact or broad keyword matching. Generally, all the ADs are ranked by the product of bid and predicted \emph{Click Through Rate (CTR)}. Once clicked, they will be charged the minimal price to maintain their position (Generalized Second Price, GSP), to maximize the expected revenue of the platform \cite{King07}. Cumulative post-click behaviors such as adding to collections/shopping carts and making purchases are used as ranking signals in future organic search, in which way TSA is especially useful for the cold start of new arrivals and earning persistent revenue.

% Problem and main challenges
Apart from keyword selection, advertisers can optimize their marketing demands in TSA by way of cautiously determining strategic bids. Nevertheless, such optimization is intractable due to the shortage of auction competition information, computation, and storage capacity. On the contrary, the advertising platform is omniscient and amenable to help advertisers bid for a healthy marketing ecology. However, if the platform takes the duty to design a bidding strategy for advertisers, new challenges might appear. Firstly, under the \emph{Pay Per Click (PPC)} pricing mechanism, a mismatch of profits naturally arises between the platform's immediate revenue and advertiser's Gross Merchandise Volume (GMV), i.e., the platform charges the click while most advertisers bid for the conversion/purchase. Secondly, any bidding strategy designed by the platform is inevitable to be faced with the multi-agent problem, i.e., an uncertain amount of heterogeneous advertisers might opt in proxy bidding, which makes the strategy difficult to optimize. Last but not least, millions of varied-sized advertisers hosted by TSA have quite different marketing demands, showing diverse preferences in impression-level bidding. Existing work mostly falls into the maximization of a single advertiser's profit in display advertising by the Demand Side Platforms (DSPs, e.g., iPinYou \cite{Ren17,Zhang17,Zhang14}, JD \cite{Wang17}, and Media6Degrees \cite{Perlich12}), and are not motivated to take platform revenue and consumer experience into account, which is critical for an advertising platform.

% Contribution of this paper.
In this paper, we cater to various advertiser demands with a unified impression-level bidding based demand optimization framework called \emph{Customer Intelligent Agent (CIA)}. CIA keeps advertisers' knobs on keyword-level bids to convey varied-level Return-On-Investment (ROI) expectations. Meanwhile, the impression-level bid is applied considering both real-time AD-query features and advertisers' demand-specific ROI preferences. Combined with CTR and \emph{CVR (Conversion Rate)} prediction modules \cite{Gai17CTR}, CIA balances the benefits of three parties without changing the original auction mechanism. A replay based simulation system takes the historical auction in to evaluate the winning probability and daily performance of different bids as the optimization tool. Offline simulations and online A/B test show that it outperforms optimized fixed keyword-level bids for various demands fulfilling. CIA based GMV optimization and style comparison have been widely adopted by advertisers \footnote{More than $50\%$ of all the ADs switched to CIA bidding, contributing over $30\%$ of the daily revenue according to the statistics in Feb. 2018.} in the TSA platform.

% Organization of this paper
The rest of this paper is organized as follows. In Section \ref{sec:related_work} we give a more detailed analysis of related work on bidding optimization. The architecture of CIA is given in Section \ref{sec:architecture} where we formulate the optimization problems with their constraints. Section \ref{sec:ranking} discusses the influence of CIA bidding in the GSP mechanism. And we give a replay based implementation of CIA in Section \ref{sec:replay}. In Section \ref{sec:experiments} we report some simulation and experimental results illustrating the performance of CIA on the TSA system. Section \ref{sec:conclusion} concludes the paper.

\section{Related work}\label{sec:related_work}
Most existing bidding strategies designed by DSPs aim to optimize only one single advertiser's profits by assuming that the competitive environment is stationary \cite{Diemert17,Ren17,Zhang14,Perlich12}. In the case of unconstrained budget, truthful bidding in GSP has proved to achieve the Nash-Equilibrium for a single item auction \cite{Krishna09}. In \cite{Zhang14}, Zhang et al. studied the optimal bidding in display advertising with budget constraints on sequential bids, with the expected number of clicks maximized. Relying on independent CTR and bid landscape prediction modules, the authors proved that concave bidders paying more attention to low CTR impressions might be better. Later, if the budget was directly on the charging price, Zhang et al. proved that linear bidding is optimal \cite{Ren17} and a joint optimization including CTR estimation, market price prediction and bidding were applied. However, all these methods consider only a single advertiser's profits. Moreover, a reasonable budget constraint is usually not directly available in TSA practice.

Other novel work also exists in display advertising. Considering the sequential decision processes in impression bidding, Cai et al. utilized reinforcement learning to bid \cite{Zhang17}. With the assumption of i.i.d impression features, they derived an optimal real-time bidder under budget and auction volume constraints. A similar idea was also used in JD DSP business, with deep Q-learner in its bidder\cite{Wang17}. Reinforcement learning is a straightforward solution to the scenario of delayed rewards in online advertising. Diemert et al. improved single-impression bidder by attributing post-impression actions into previous bids to decouple the sequential bidding process \cite{Diemert17}. Another related line of bidding optimization research lies in smart pacing, which helps budget-limited advertisers decide whether to bid or an impression or not \cite{Karande13,Lee13}.

In sponsored search, the work on joint optimization of keyword-level bids and budget allocation over a campaign (several campaigns shared a budget) \cite{Zhang12} resembles our scenario. More keyword-level sponsored search bidding work can be found in \cite{Liu15}. However, keyword-level bidding does not distinguish impressions from the same query. Such coarse-grained bidding is incapable of serving advertisers in an optimal way or meeting the ROI constraints well.

\section{CIA Framework in Taobao Sponsored Search}\label{sec:architecture}
% CIA architecture description
Marketing demands at TSA can be classified into two categories, i.e., basic and compound demands. Basic demands include acquisitions of impressions, clicks and conversions, which are ubiquitous in online advertising. Particularly, for e-commerce platforms like Taobao, compound demands are more platform-specific, which cover the complete marketing cycle of an AD. Detailed demands and their Key-Performance-Indicators (KPIs) are summarized in Tab. \ref{tab:demands}.
\begin{table}[htp]
\tiny
\caption{Various advertiser demands and KPIs in TSA}\label{tab:demands}
\begin{tabular}{c|c|c}
\hline
Demands & Description & KPI \\
\hline \hline
\multirow{3}{*}{Basic} & Impression & Number of impressions \\
\cline{2-3}
~ & Click & Number of clicks \\
\cline{2-3}
~ & Conversion & Collections/Shopping carts/GMV \\
\hline
\multirow{3}{*}{Compound} & Style comparison & Fast comparison of different styles under a budget \\
\cline{2-3}
~ & Audience expansion & GMV from new consumers \\
\cline{2-3}
~ & Price-off promotions & Fast clearance with an acceptable ROI \\
\hline
\end{tabular}
\end{table}

\begin{figure*}[!htb]
\centering
\includegraphics[width=6.0in]{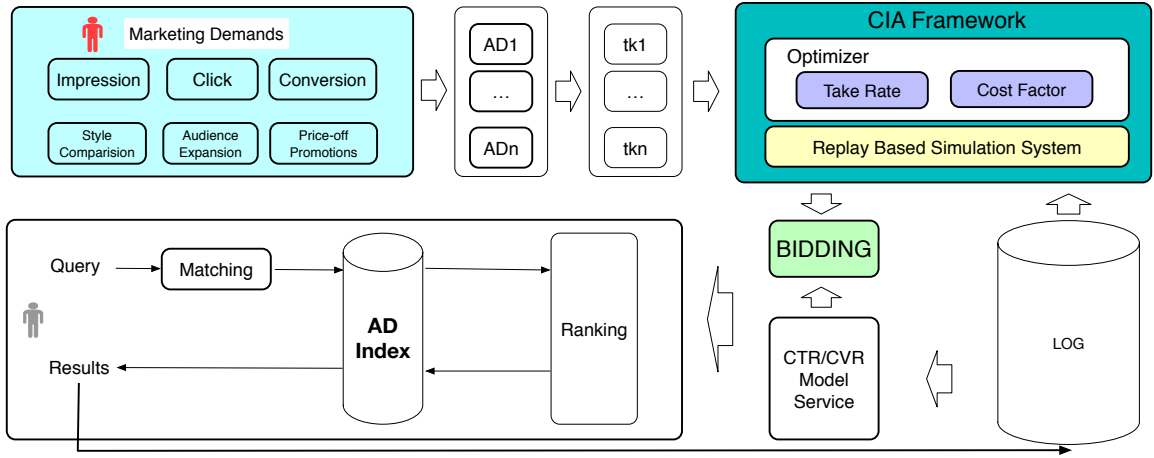}
\caption{Typical query processing procedure via CIA within TSA.}\label{fig:system_architecture}.
\end{figure*}

Fig. \ref{fig:system_architecture} illustrates how CIA framework works within the TSA system. Advertisers express their marketing demands on different ADs via keyword-level bids or direct AD-level take-rate (abbreviated as $tk$) and CIA optimizes impression-level bids respecting the demands. Optimizer balances the take-rate and advertising costs. A replay based simulation system is utilized to obtain the solution. CTR/CVR prediction models are also necessary to characterize an impression but will not be elaborated in detail as they are not the key contributions of our work. All the competitive ADs and their contextual information in an auction are recorded in logs for the replay system.

% From optimization section to the architecture section
Advertisers' demands are optimized under CIA in equation (\ref{eqn:general_optimization}). Regardless of various demands, a reasonable budget constraint is necessary. Additionally, advertisers also bid with the expectation of conversions (adding to collections/shopping carts/purchase) in e-commerce. Each dollar spent should account for some returns, only high or low for different demands. Therefore, demand optimization should also take unit conversion cost (aka $tk$) as a constraint.
\begin{equation}\label{eqn:general_optimization}
\begin{aligned}
\max\limits_{bid} \quad & \text{KPI} \\
s.t. \quad & \text{budget constraint}, \\
& \text{unit conversion cost constraint}.
\end{aligned}
\end{equation}

While optimization objectives are exactly demonstrated in Tab. \ref{tab:demands}, the constraints, despite roughly depicted above, are still challenging to be specified. Firstly, budget constraints are set only at the campaign level in TSA, while AD-level $tk$ to regulate impression-level bidding requires a reasonable AD-level budget. Secondly, advertisers prefer to use fixed keyword-level bids instead of $tk$ to express their expected unit conversion cost due to business secrets or inaccurate CTR predictions. To sum up, we have to infer reasonable $tk$ and also budget from keyword-level bid settings. We will elaborate on the inference in the following section.

\subsection{Bridging keyword-level bid settings and constraints}\label{sec:bidding}
Generally, advertisers choose the keywords of an AD and set fixed bids. Given $m$ auctions ($j=1,...,m$) which involves the AD, the accumulated ROI can be calculated as:
\begin{equation}\label{eqn:adroi}
R = \frac{\sum\limits_{j=1}^{m}{ctr_j * cvr_j * ip_j}}{\sum\limits_{j=1}^{m}{ctr_j * c_j}} \approx \frac{\sum\limits_{j=1}^{m}{ctr_j * cvr_j * ip_j}}{\sum\limits_{j=1}^{m}{ctr_j * b_j}},
\end{equation}
where $c_j$ is the click cost and can be approximated with the corresponding keyword bid $b_j$ to obtain a lower bound of ROI while $ip$ stands for the item price. Following the above formula, we actually obtain an expected ROI from the original bid settings and auction log. Implicitly, this ROI reflects the advertiser's implicit rationality in the marketing environment.

Meanwhile, it is sufficient to guarantee the cumulative ROI if each advertiser bids with an acceptable ROI in each auction. An impression-level $roi$ satisfies that,
\begin{equation}\label{eqn:pvroi}
roi = \frac{ctr * cvr * ip}{ctr * c},
\end{equation}
where $c$ is the cost upon clicking. Let $roi \ge R$, then $c \le \frac{cvr*ip}{R}$. In GSP it usually satisfies that $c \le bid$, therefore it is sufficient to bid as,
\begin{equation}\label{eqn:tkbidding}
bid = tk * cvr * ip, \quad tk = \frac{1}{R}.
\end{equation}
Furthermore, the advertiser's changing keyword-level bids can easily propagate to $tk$ given the historical auction log.
\begin{equation}
\Delta tk \approx \frac{\sum\limits_{j=1}^{m}{ctr_j * \Delta b_j}}{\sum\limits_{j=1}^{m}{ctr_j * cvr_j * ip_j}},
\end{equation}
where $\Delta(\cdot)$ corresponds to the perturbations on its input argument. 

In the above analysis, we take the auction log of an AD, arriving an AD-level $tk$. Campaign/keyword-level $tk$s can be derived likewise with corresponding logs. $tk$ can be updated daily with historical auction logs or in an online way from real-time auctions. In the latter way, real-time changes in advertisers' keyword-level bid settings and auction distribution can be easily captured. In this paper, we take the AD-level $tk$ for the daily distribution of an AD's auctions is relatively stationary (which will be justified in Section \ref{sec:replay}), and also for its high-interpretability to advertisers.

Relying on the historical auction log, a reasonable AD-level virtual budget can also be obtained as:
\begin{equation}\label{eqn:budget}
B = \sum\limits_{j=1}^{m}{ctr_j * b_j}.
\end{equation}
If the marketing environment is stationary across days, we believe that such a budget shall be fair and sufficient for the advertiser. In this way, we do not need to explicitly split the campaign budget into each AD.

To optimize various campaign-level demands, CIA introduces an additional \emph{cost regulation factor} $\alpha$ to adjust AD-level $tk$ in a campaign, leading to a generalized bidding formula:
\begin{equation}\label{eqn:bidcf}
bid = \alpha * tk * cvr * ip.
\end{equation}
Generally, an $\alpha$ in the range of $(0,1]$ can roughly guarantee that the advertiser will gain a higher ROI than previous keyword-level bidding. However, for the approximations used, the de facto effective $\alpha$ might be larger than $1$. Additionally, different demands might lead to a quite different feasible range of $\alpha$. For each AD in a campaign, query distribution is quite different from each other but shares a common budget. For various demands, this leaves us space to optimize different objectives by regulating AD-level $\alpha$.

\subsection{Demands optimization}\label{sec:demands_optimization}
An AD-level $tk$ simplifies advertisers' bidding control. Meanwhile, campaign-level demand is the most natural way for advertisers to manage. To this end, in this section we tackle the two most representative demands of advertisers, i.e., campaign-level GMV (conversion) optimization and style comparison by regulating $AD$-level $tk$ (aka, $\alpha$) to illustrate the whole optimization process of CIA framework. 

% CPC range for knapsack, demands leads to cost, thus ROI.
In the following formulation, for a campaign with $n$ ADs (indexed by $i=1,...n$), we denote the expected daily GMV and cost of ADs if bidding with CIA as $\mathbf{y}$, $\mathbf{z} \in \mathbb{R}^n$. Under the same auction environment, corresponding keyword-level bids will lead to a cost of $\tilde{\mathbf{z}}$. And the cost regulation vector is denoted as $\bm{\alpha} \in \mathbb{R}_{+}^n$. Beyond standard keyword-level bids, advertiser's bids of the $n$ ADs lies in the range of $[\mathbf{l}, \mathbf{u}]$, $\mathbf{l}, \mathbf{u} \in \mathbb{R_{+}}^n$. $\mathbf{1}$ represents the $n$-dimensional column vector with all its elements to be one. $\alpha_i$ is the cost regulation factor of the $tk$ for the $i$-th AD. The same subscript logic applies to $\mathbf{y}$ and $\mathbf{z}$. $\mathbf{x}^T$ denotes the transpose of $\mathbf{x}$. $||\cdot||_2$ is the Euclidean norm of the input argument.

\subsubsection{Advertiser campaign GMV optimization}
Campaign-level GMV optimization is fundamental for advertisers in TSA. For a campaign, we want to determine each AD's $\alpha_i$ so that GMV is maximized. For a campaign-level GMV, the problem can be formulated as:
\begin{equation}\label{eqn:gmvelevation}
\begin{aligned}
\max\limits_{\bm{\alpha}} \quad & \mathbf{1}^T \mathbf{y} \\
s.t.\quad & (\beta - \epsilon) \mathbf{1}^T \tilde{\mathbf{z}} \le \mathbf{1}^T \mathbf{z} \le (\beta + \epsilon) \mathbf{1}^T \tilde{\mathbf{z}}, \\
\quad & \beta \in (0,1], \epsilon \in [0,\beta),
\end{aligned}
\end{equation} 
where $\beta$ is the cost regulation parameter and $\epsilon$ is the permissible offset determined from the demands. In practice, $\beta$ is set no larger than $1$ to ensure that the cost does not exceed the keyword bidding cost, i.e., the virtual budget. Meanwhile $\bm{\alpha}$ is constrained in ranges determined from advertiser's tolerable ranges of $[\mathbf{l}, \mathbf{u}]$.

CIA generates fine-grained impression-level bids with coarse-grained AD-level $\bm{\alpha}$ and $tk$ for implicit ROI constraints. Note that in basic CIA, $\alpha_i \le 1$ is used to guarantee a higher ROI while such constraint on $\alpha_i$ is eliminated but replaced with the ROI range implied by $[l_i, u_i]$ in the GMV optimization. To solve the problem, we first utilize the bid ranges $[\mathbf{l},\mathbf{u}]$ to derive a feasible range of each $\alpha_i$. For this purpose, firstly $l_i$ and $u_i$ are replayed as keyword-level bids to obtain the corresponding cost range of the $i$-th AD. This is reasonable since the cost is monotonically non-increasing with the bids. Then the replay module is inversely applied to approximate the feasible range of $\alpha_i$ from cost.  

Assume there is a set of $k$ effective $\alpha_i$, denoted as $\{\alpha_{i1},...,\alpha_{ik}\}$ \footnote{For the ease of presentation, we assume that $k$ is the same among all ADs. In practice $k$ could be varied due to bid settings of different ADs.}. For each AD, $\alpha_{ij} (j=1,...,k)$ is evaluated by the replay module to yield a valuation point $(\alpha_{ij}, y_{ij}, z_{ij})$. $\tilde{z}_i$ is obtained by replaying with keyword-level bids. Following the discretization, the GMV optimization problem can be reformulated as follows: 
\begin{equation}\label{eqn:gmvknapsack}
\begin{aligned}
\max\limits_{\{x_{ij}\}} \quad & \sum\limits_{i=1}^{n}{\sum\limits_{j=1}^{k}{x_{ij} y_{ij}}} \\
s.t. \quad & \sum\limits_{j=1}^{k}{x_{ij}} = 1, x_{ij} \in \{0,1\}, \forall i = 1,...,n, \\
\quad & (\beta - \epsilon) \sum\limits_{i=1}^{n}{\tilde{z}_i} \le \sum\limits_{i=1}^{n}{\sum\limits_{j=1}^{k}{x_{ij}z_{ij}}} \le (\beta + \epsilon) \sum\limits_{i=1}^{n}{\tilde{z}_i}
\end{aligned} 
\end{equation}
where $x_{ij}$ is an indicator variable that $x_{ij} = 1$ means $\alpha_{ij}$ is selected for the $i$-th AD. The above formulation belongs to a classic dynamic programming problem and can be solved with group Knapsack algorithm \cite{Salkin1975knapsack}. After obtaining the optimal $\{x_{ij}\}$, optimal $\bm{\alpha}$ can be retrieved accordingly.

\subsubsection{Style comparison}
For an FMCG store, the launching of new clothing styles now and then is common practice. To reduce the marketing risk of awful designs, advertisers need to be aware of the market response to different styles quickly to adjust future production and inventory. Within CIA, different fashion styles should consume nearly equal impressions to compare their post-impression performance in a quantifiable way. Advertisers only need to put the $n$ tested styles in the same campaign and let CIA take over. Given a comparable budget with keyword bidding, CIA solves the following optimization problem to allocate impressions to each style as uniform as possible.
\begin{equation}
\begin{aligned}
\max\limits_{\bm{\alpha}} \quad & -||\mathbf{s} - \frac{\mathbf{1}^T \mathbf{s}}{n}\mathbf{1}||_2 \\
s.t. \quad & (\beta - \epsilon) \mathbf{1}^T \tilde{\mathbf{z}} \le \mathbf{1}^T \mathbf{z} \le (\beta + \epsilon) \mathbf{1}^T \tilde{\mathbf{z}}, \\
\quad & \beta \in (0,1], \epsilon \in [0,\beta),
\end{aligned}
\end{equation}
where $\beta$ and $\epsilon$ have the same meaning as in GMV optimization and the new variable $\mathbf{s}$ denotes the amount of impressions in all the ADs.

This problem can also be solved by group Knapsack algorithm following the same logic with GMV optimization. However, for the quadratic form of the objective, the optimal solution would yield uniform impressions among ADs. Based on this, we first calculate the feasible cost range as $[b,B]$. Then $l_i$ and $u_i$ are replayed to obtain a feasible range of $\alpha_i$ and also $[b_i, B_i]$. Following the notations, the style comparison problem can be reformulated as:
\begin{equation}
\begin{aligned}
\max\limits_{\bm{\alpha}} \quad & -||\mathbf{s} - \frac{\mathbf{1}^T \mathbf{s}}{n}\mathbf{1}||_2 \\
s.t. \quad & b_i \le z_i \le B_i, \forall i = 1,...,n, \\
\quad & b \le \mathbf{1}^T \mathbf{z} \le B.
\end{aligned}
\end{equation}

The above formulation is a standard quadratic programming problem and can be solved with mature solvers such as OSQP \cite{osqp}. After obtaining the optimal $\mathbf{s}$, the replay module combined with the binary search method is utilized to get the optimal $\bm{\alpha}$. Compared with Knapsack algorithms, directly optimizing with quadratic programming provides a numeric solution without manual discretization.

% Ranking should be explained with additional analysis on mechanism level.
\section{Ranking with CIA}\label{sec:ranking}
% Two main parts: 1. unifying three party profits in the ranking formula; 2. influence on the mechanism for multi-agent behavior.
Within keyword bidding, there exists an obvious mismatch: platform tries to optimize $ctr$ for revenue while advertisers care more about $cvr$. Meanwhile, consumer experience depending on conversions (consistent with advertisers) is also ignored. To mitigate the mismatch, industrial practice often introduces all kinds of relevant factors into the ranking formula (such as audience/advertiser experience) \cite{Zhu17} or implements a multi-stage ranking with different criteria \cite{Liu2017} to balance advertisers' profits. Nevertheless, all these strategies bring damage to the classic auction mechanism, affecting its interpretation and the platform's revenue.

Given the new bidding formula, the ranking score in CIA has the following form:
\begin{equation}\label{eqn:ranking}
r = bid * ctr = \alpha * tk * cvr * ip * ctr = \alpha * tk * gmv.
\end{equation}
Compared with the original ranking score, CIA takes explicitly $cvr$ into bids while the ranking formula is preserved. In essence, budget savings from low-quality impressions are spent on higher-valued impressions, leading to a higher ROI. Therefore, the advertiser's GMV increases together with platform revenue and audience experience. 

We argue that the CIA bidding keeps the GSP auction mechanism unchanged, i.e., the allocation and charging mechanisms have remained the same. Advertisers can still change their keyword-level bid settings to influence $tk$ so that the auction game still continues. For multi-agent behavior, the CIA bidding only takes individual $tk$ and impression value into account, therefore bypasses the difficulty of modeling the competition's uncertainty to a great extent. However, compared with keyword-level bidding, CIA bids on an impression-level and bases its decision on conversion value (which is typically unavailable for advertisers), thus helping alleviate advertisers' irrationality and might shorten the time to reach Nash equilibrium in the repeated auction game. In this sense, the market is more efficient.

\section{Replay based CIA implementation}\label{sec:replay}
To support optimization in the CIA framework, we need to predict the performance of different $\alpha$ on an AD. Particularly, we are interested in the following two functions:
\begin{equation*}
\begin{aligned}
& f(\alpha, AD, \mathcal{E}): \text{daily cost of an AD if bidding based on $\alpha$}, \\
& g(\alpha, AD, \mathcal{E}): \text{daily GMV of an AD if bidding based on $\alpha$},
\end{aligned}
\end{equation*}
where $\mathcal{E}$ represents the external bidding environment. For the baseline cost, we have:
\begin{equation*}
h(KB, AD, \mathcal{E}): \text{daily cost of an AD with keyword bidding}.
\end{equation*}

\begin{figure}[!htb]
\centering
\includegraphics[width=2.2in]{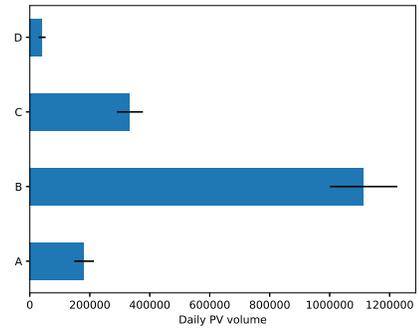}
\caption{Auction volumes of four ADs across a week, where the length of the rectangular displays the mean and the line shows the standard deviation (A-Home Fabric, B-Cosmetics, C-Female Clothing, D-Office Furniture).}\label{fig:PV_volume}
\end{figure}

\begin{figure*}
  \centering
  \subfigure[AD-A]{
    \label{fig:AD-482135539} %% label for first subfigure
    \includegraphics[width=1.65in]{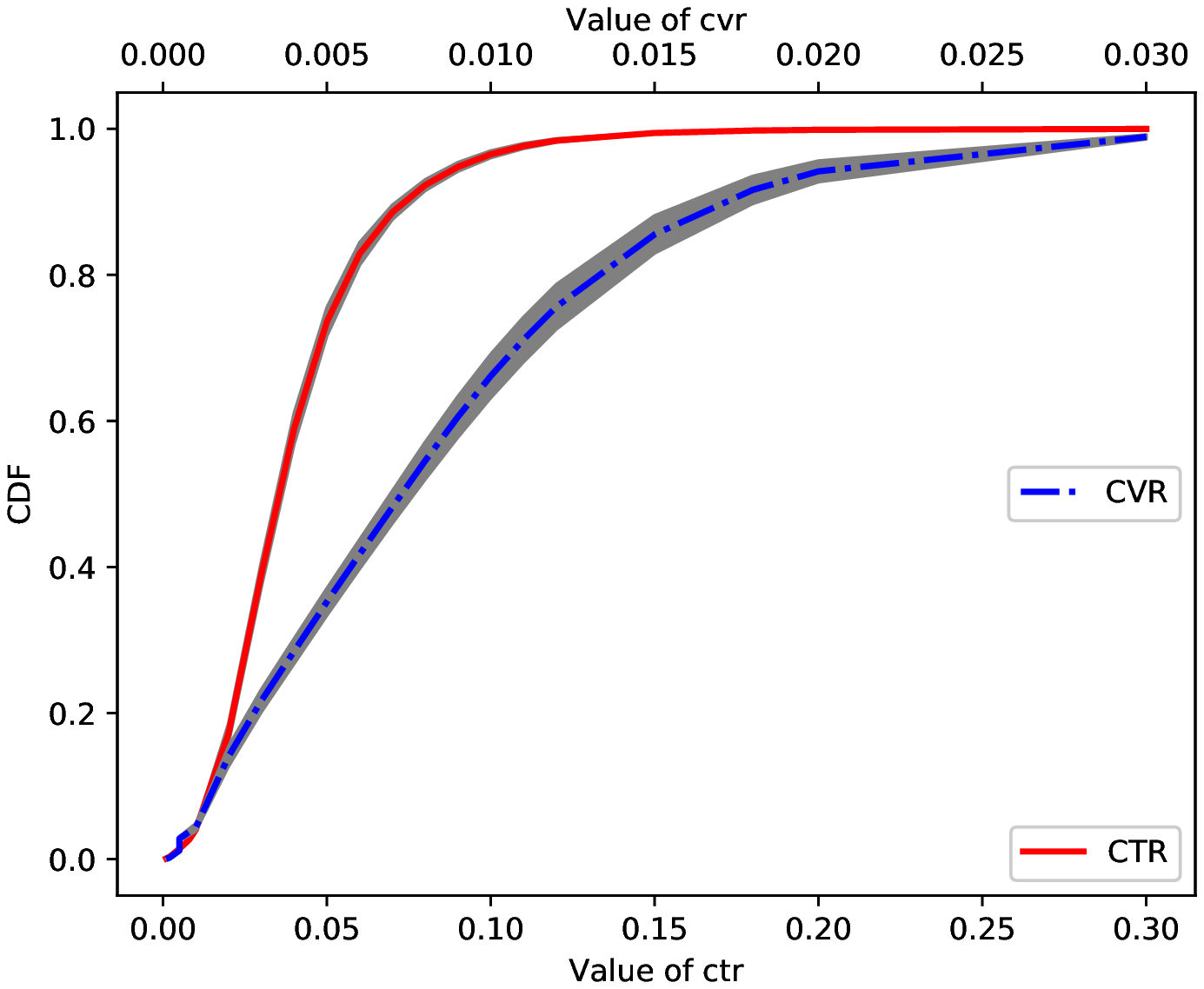}}
    %\hspace{3.5em}
  \subfigure[AD-B]{
    \label{fig:AD-485704514} %% label for second subfigure
    \includegraphics[width=1.65in]{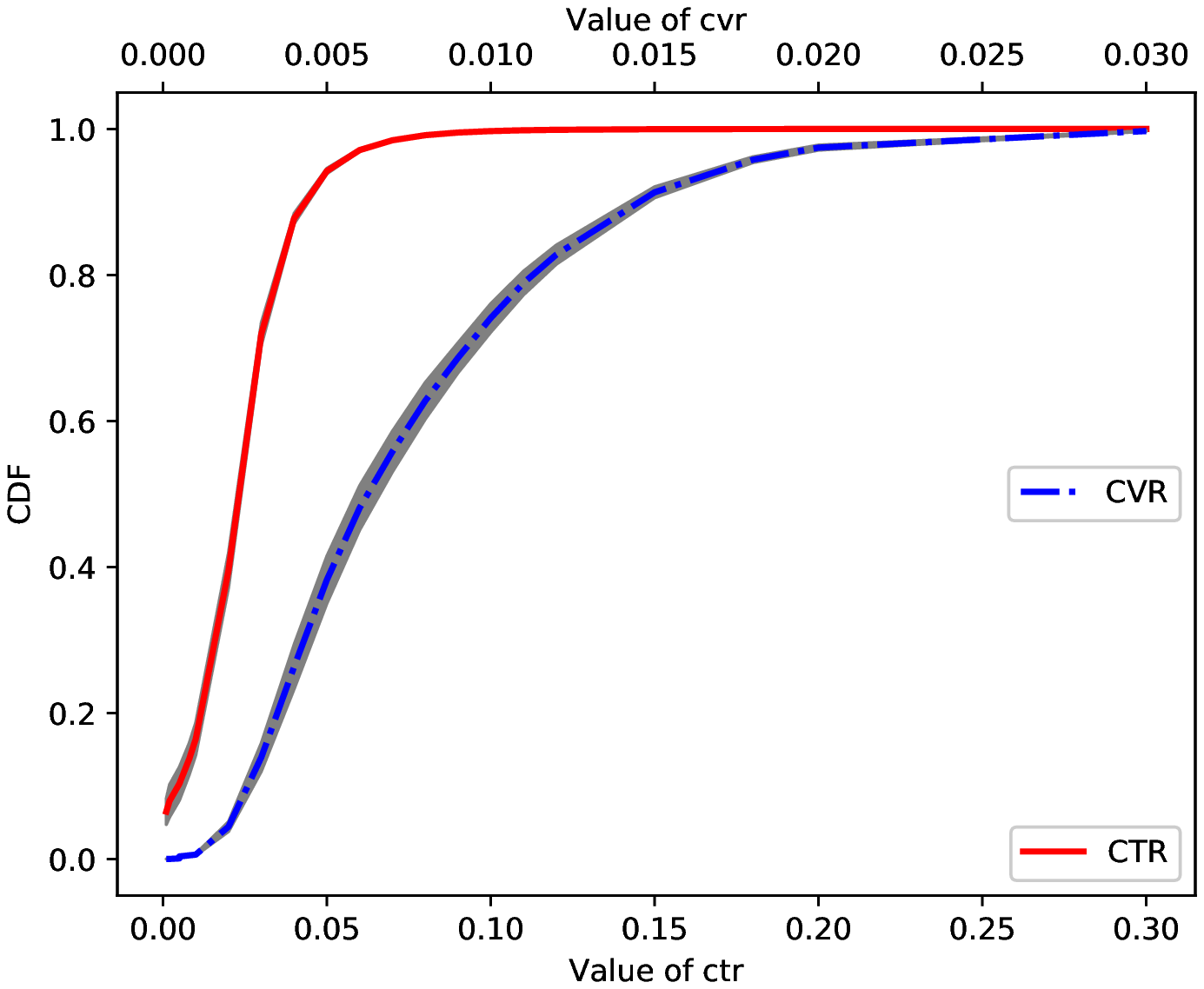}}
    %\hspace{3.5em}
  \subfigure[AD-C]{
    \label{fig:AD-783731892} %% label for second subfigure
    \includegraphics[width=1.65in]{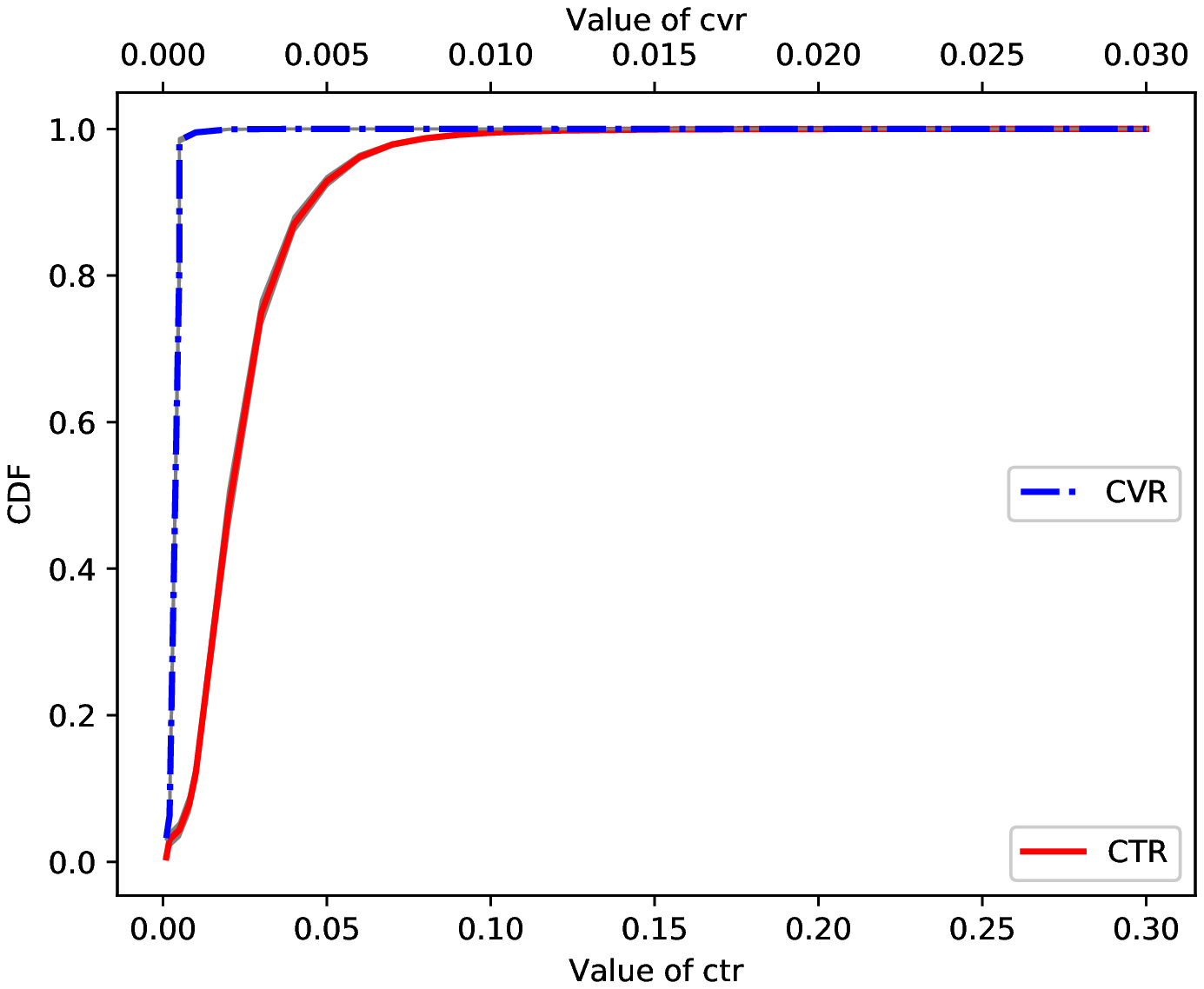}}
    %\hspace{3.5em}
  \subfigure[AD-D]{
    \label{fig:AD-783739714} %% label for second subfigure
    \includegraphics[width=1.65in]{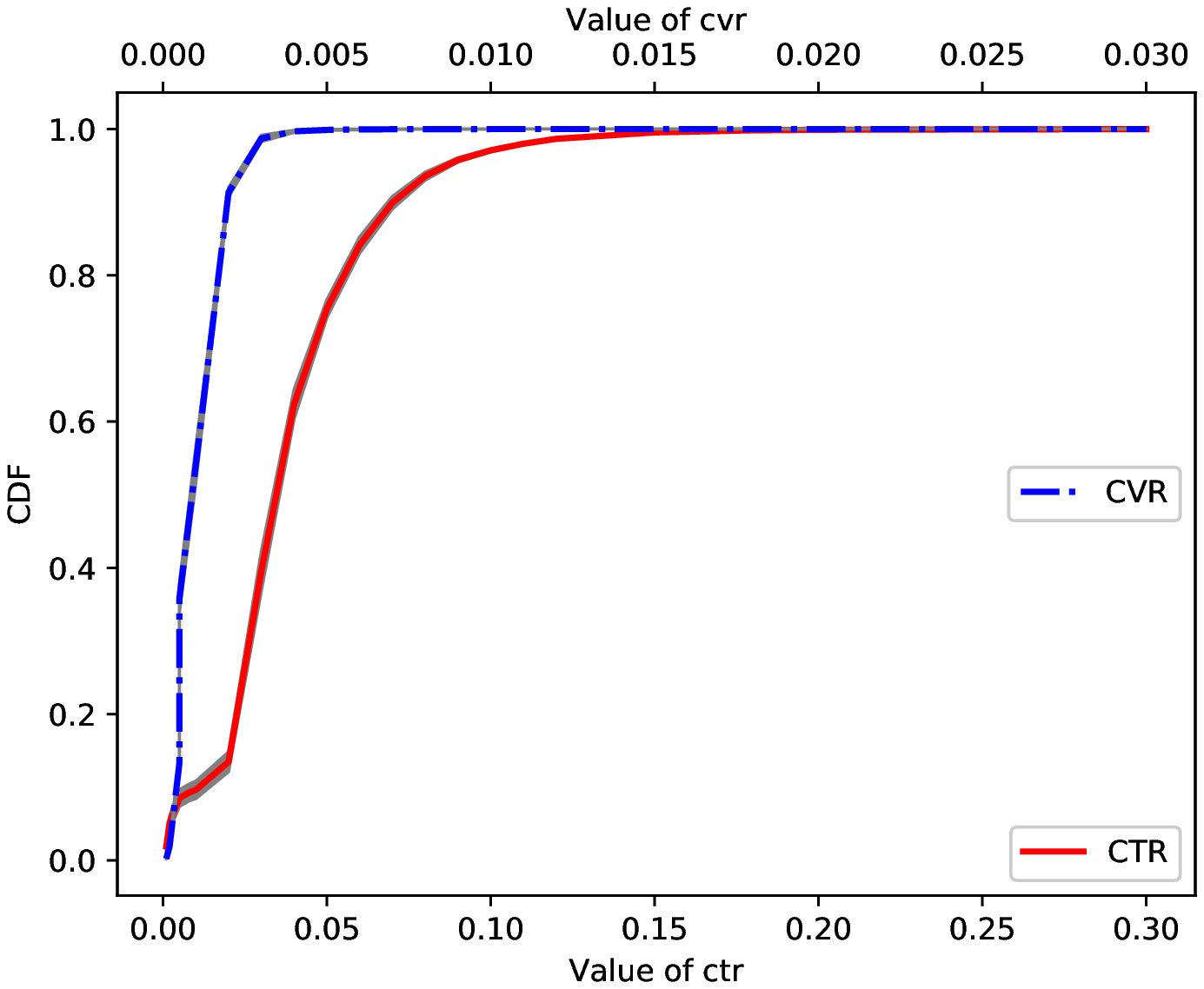}}
  % \vspace{-1.5em}
  \caption{CTR and CVR distributions of four ADs' auctions across a week. The line represents the mean while the shadow represents the standard deviation.}
  \label{fig:distributions} %% label for entire figure
\end{figure*}

A historical auction log contains the contextual information of all the candidate ADs, bids, predicted $ctr$, $cvr$ and commodity price etc. An asynchronous data collection system based on MaxCompute \cite{MaxCompute} is designed to tackle the huge-volume streaming queries. For the replay-based prediction to be usable, the auction distribution of each AD should be relatively stationary. To justify the assumption, we chose four typical ADs to plot their daily auction statistics from Jan. 25, 2018 to Jan. 31, 2018 in Fig. \ref{fig:PV_volume}. Furthermore, the corresponding cumulative CTR and CVR distributions are displayed in Fig. \ref{fig:distributions}, which shows that CTR and CVR distributions are nearly the same for all the test ADs across the week. The two results together justify that the replay module is reliable enough to predict performance based on historical auction records.

With the historical daily auction log, the replay works as:
\begin{enumerate}
	\item Updating the bids with $\alpha$, and re-ranking all the ADs in the current auction according to eCPM;
	\item For the top-$N$ ADs \footnote{$N$ is the number of ad slots for a query.}, calculating their cost $c$ once clicked;
	\item Accumulating $ctr*c$ and $ctr*cvr*ip$ to the cost and GMV, respectively.
\end{enumerate}
Keyword-level bid settings can also yield rough performance following the same procedure. Reversely, in the optimization process, a mapping from the AD's daily cost to $\alpha$ would also be necessary. The inverse estimation of $\alpha$ can be solved efficiently via binary search due to monotonicity.

As a black box module, replay encapsulates the functions of winning probability/bid landscape estimation and bidding optimization together (see Fig. \ref{fig:system_architecture}). Currently, it can generate a pair of $(\alpha, cost/GMV)$ in milliseconds. To alleviate the influence of marketing shifts, CIA can average performance from the auction log of the past several days. Other parameterized representations can also fit the mappings (see \cite{Ren17} and references therein). However, we choose to replay for its time efficiency and validated accuracy in engineering practice. 

\section{Experimental Evaluation}\label{sec:experiments}
In this section, we first introduce our experimental settings including baseline algorithms and evaluation metrics. Then we report some offline simulation and online A/B test results with further analysis and discussions. 

\subsection{Experimental settings}
To test the performance of the CIA framework, we decompose evaluations into three levels, i.e., AD level, campaign level and platform level.

The AD level evaluation bids as equation (\ref{eqn:bidcf}) (abbreviated as CIA-AD) and is compared with keyword-level bids. The evaluation of CIA-AD is only used to illustrate the effectiveness of CIA over keyword-level bids. For the AD-level evaluation, we compare the performance of GMV, ROI, CVR, and PPC. Formally, CIA-AD is evaluated with the following baseline with approximately the same advertising cost ($\beta = 1, \epsilon = 0.1$):\\
\textbf{KB:} Standard keyword-level bid settings are used to bid each impression.\\

Campaign level experiments constitute the core evaluations. We test the GMV optimization (CIA-CAMP1) and style comparison (CIA-CAMP2) performance in Section \ref{sec:demands_optimization}. The following baselines are used ($\beta = 1, \epsilon = 0.2$):\\
\textbf{CoKB:} For campaign GMV optimization, joint optimization of budget allocation and keyword-level bids is implemented \cite{Zhang12}, to compare with CIA-CAMP1.\\
\textbf{KB-CAMP:} For style comparison, advertisers set the same keyword bids on different ADs to compare their performance, which serves as the baseline of CIA-CAMP2. The normalized standard deviation of cost spent on different ADs in CIA-CAMP2 is compared.\\ 

The platform level experiments mainly focus on evaluating the effects on the platform when massive advertisers switch to CIA bidding. Specifically, advertiser GMV/ROI and platform revenue are evaluated simultaneously. 

To make the results more persuasive, all the experiments are carried out on massive randomly selected ADs/campaigns across two weeks. For the sake of commercial privacy, all performance measures are presented in the form of a relative shifting ratio compared with the baseline.

\subsection{Offline simulation}
The offline simulation operates on historical auction logs and relies on replay for optimization and evaluation.

\textbf{AD-level:} We randomly choose $133,743$ ADs and test their performance with CIA-AD. The overall average performance and four examples are shown in Tab. \ref{tab:CIA-AD-offline}
\begin{table}[htp]
\centering
\tiny
\caption{Offline CIA-AD performance with KB as a baseline}\label{tab:CIA-AD-offline}
\begin{tabular}{|c|c|c|c|c|c|c|}
\hline
Measures (\%) & cost & GMV & ROI & CVR & PPC \\
\hline \hline
\textbf{Overall} & $\mathbf{-0.10}$ & $\mathbf{9.69}$ & $\mathbf{9.86}$ & $\mathbf{10.90}$ & $\mathbf{1.80}$ \\
\hline
AD-A & 0.90 & 6.00 & 5.10 & 6.20 & 1.00 \\
\hline
AD-B & -0.09 & 9.54 & 9.64 & 6.03 & 4.66 \\
\hline
AD-C & 0.51 & 18.84 & 18.24 & 23.27 & 4.25 \\
\hline
AD-D & 0.53 & 6.74 & 6.18 & 6.55 & 0.35 \\
\hline
\end{tabular}
\end{table}

From Tab. \ref{tab:CIA-AD-offline} we can see that with an approximately equal cost, the GMV and ROI of CIA-AD has an increase of nearly $10$ percent. Meanwhile, CVR obtains the maximum increase of $10.9$ percent, which means that more impressions with high conversion rates are acquired by CIA bidding. This coincides with our analysis as the bidding equation takes $cvr$ into account. It can also be seen that with more cost, AD-A, C, and D even see an ROI elevation, which means that platform revenue also grows with advertisers' ROIs.

\textbf{Campaign-level:} For a campaign-level offline simulation, the results of randomly selected $2599$ campaigns which opt in CIA are shown in Tab. \ref{tab:CIA-CAMP1-offline}. It shows that with a nearly equal cost, CIA leads to an increase of GMV/ROI by $13$ percent, which is reasonable since CoKB still bids on the keyword level although co-optimized. Meanwhile, CIA bids towards high-valued impressions in the most fine-grained way. The increase in CVR also validates the analysis.
\begin{table}[!htp]
\centering
\tiny
\caption{Offline CIA-CAMP1 performance with CoKB bidding as a baseline}\label{tab:CIA-CAMP1-offline}
\begin{tabular}{|c|c|c|c|c|c|c|c|}
\hline
Measures (\%) & cost & GMV & ROI & CVR & PPC \\
\hline \hline
\textbf{Overall} & $\mathbf{1.18}$ & $\mathbf{14.64}$ & $\mathbf{13.62}$ & $\mathbf{17.10}$ & $\mathbf{3.80}$ \\
\hline
CAMP-A & -0.25 & 15.70 & 15.99 & 28.41 & 11.22 \\
\hline
CAMP-B & 0.24 & 14.48 & 14.21 & 22.34 & 8.63 \\
\hline
CAMP-C & -0.87 & 6.41 & 7.34 & 18.02 & 12.07 \\
\hline
CAMP-D & 2.25 & 25.00 & 22.25 & 32.45 & 8.35 \\
\hline
\end{tabular}
% \hspace{-3em}
\end{table}

For the style comparison demand, the performance of CIA-CAMP2 with $17334$ campaigns is listed below in Tab. \ref{tab:CIA-CAMP2-offline}. Compared with naive equal keyword bids for style comparison, CIA-CAMP2 arrives a more uniform impression allocation, leading to a fair style comparison.
\begin{table}[!htp]
\centering
\tiny
\caption{Normalized standard deviations of offline style comparison performance}\label{tab:CIA-CAMP2-offline}
\begin{tabular}{|c|c|c|}
\hline
Strategies & \textbf{CIA-CAMP2} & \textbf{KB-CAMP}  \\
\hline \hline
\textbf{Overall} & $\mathbf{0.660}$ & $\mathbf{0.790}$ \\
\hline
CAMP-A & 0.520 & 0.799 \\
\hline
CAMP-B & 0.852 & 0.951 \\
\hline
CAMP-C & 0.181 & 0.269 \\
\hline
CAMP-D & 1.562 & 1.687 \\
\hline
\end{tabular}
% \hspace{-3em}
\end{table}

\textbf{Platform level:} Multiple advertisers to opt in CIA will definitely change the global bidding environment, bringing complex consequences. To further evaluate the effects, we conduct an offline performance test by randomly switching a proportion of all the ADs to bid with CIA. The results of the whole platform (including all ADs) are displayed in in Tab. \ref{tab:CIA-offline}.
\begin{table}[htp]
\centering
\tiny
\caption{Offline platform performance with KB bidding as a baseline}\label{tab:CIA-offline}
\begin{tabular}{|c|c|c|c|c|c|}
\hline
Measures (\%) & \textbf{cost} & GMV & \textbf{ROI} & CVR & PPC \\
\hline \hline
10 & $\mathbf{0.25}$ & 0.58 & $\mathbf{0.33}$ & 0.72 & 0.43 \\ 
\hline
30 & $\mathbf{0.58}$ & 3.42 & $\mathbf{2.82}$ & 2.31 & 1.00 \\
\hline
50 & $\mathbf{0.76}$ & 4.79 & $\mathbf{4.00}$ & 3.99 & 1.33 \\
\hline
100 & $\mathbf{1.41}$ & 9.08 & $\mathbf{7.5}$ & 8.18 & 1.63 \\
\hline
\end{tabular}
\end{table}
It shows that with more and more advertisers changing to CIA, the platform revenue (reflected by advertisers' costs) and advertiser ROI grows together, which means that CIA has a positive effect on a healthy marketing ecosystem. 

\subsection{Online experiments}
The online experiments compare different strategies by randomly allocating online queries to buckets implementing different strategies.

\textbf{AD-level:} We apply the replay module on the historical log to estimate a proper $\alpha$ to balance the cost. Then it is used online to evaluate the performance of CIA-AD as displayed in Tab.\ref{tab:CIA-AD-online}, where``overall'' summarizes the whole $213,510$ ADs and A,B,C,D are the same with those in offline evaluations.
\begin{table}[htp]
\centering
\tiny
\caption{Online CIA-AD performance with KB as a baseline}\label{tab:CIA-AD-online}
\begin{tabular}{|c|c|c|c|c|c|c|c|c|}
\hline
Measures (\%) & cost & GMV & ROI & CVR & PPC \\
\hline \hline
\textbf{Overall} & $\mathbf{2.00}$ & $\mathbf{11.60}$ & $\mathbf{9.50}$ & $\mathbf{11.90}$ & $\mathbf{2.30}$ \\
\hline
AD-A & -0.20 & 1.60 & 1.80 & 3.10 & 2.90 \\
\hline
AD-B & 8.80 & 12.80 & 19.10 & 2.07 & -12.90 \\
\hline
AD-C & 7.90 & 25.30 & 16.10 & 41.40 & 1.70 \\
\hline
AD-D & 6.20 & 23.60 & 16.40 & 23.00 & 7.00 \\
\hline
\end{tabular}
% \hspace{-3em}
\end{table}
With a $2$ percent increase in cost, overall GMV and ROI encounters an increase of $11.6$ and $9.5$ percent, respectively. The PPC also sees an increase of $7.0$ percent, which means that platform efficiency also increases. 

\textbf{Campaign-level:} Online evaluation results of CIA-CAMP1 and CIA-CAMP2 with the same set of offline campaigns are listed in Tab. \ref{tab:CIA-CAMP1-online} and Tab .\ref{tab:CIA-CAMP2-online}, respectively. The fundamental increase of GMV, ROI, CVR and decrease of PPC follows the right logic. We can also see that in online experiments, different campaigns have various heterogeneous performances. This might be caused by the complex environmental shift due to bidding changes. Moreover, we conclude that offline simulation results are also accurate compared with the online version, which validates that our replay system is also reliable.
\begin{table}[!htp]
\centering
\tiny
\caption{Online CIA-CAMP1 performance with CoKB bidding as a baseline}\label{tab:CIA-CAMP1-online}
\begin{tabular}{|c|c|c|c|c|c|c|c|}
\hline
Measures (\%) & cost & GMV & ROI & CVR & PPC \\
\hline \hline
\textbf{Overall} & $\mathbf{0.01}$ & $\mathbf{12.50}$ & $\mathbf{11.50}$ & $\mathbf{3.70}$ & $\mathbf{-6.29}$ \\
\hline
CAMP-A & 3.18 & 28.67 & 24.68 & 17.47 & 4.66 \\
\hline
CAMP-B & -1.58 & 43.99 & 46.32 & 4.35 & -7.07 \\
\hline
CAMP-C & -9.02 & -10.61 & -1.73 & 8.83 & -1.71 \\
\hline
CAMP-D & 6.91 & 20.37 & 12.59 & 12.56 & -4.73 \\
\hline
\end{tabular}
% \hspace{-3em}
\end{table}
\begin{table}[!htp]
\centering
\tiny
\caption{Normalized standard deviations of online style comparison performance}\label{tab:CIA-CAMP2-online}
\begin{tabular}{|c|c|c|}
\hline
Strategies & \textbf{CIA-CAMP2} & \textbf{KB-CAMP}  \\
\hline \hline
\textbf{Overall} & $\mathbf{0.524}$ & $\mathbf{0.574}$ \\
\hline
CAMP-A & 0.670 & 1.180 \\
\hline
CAMP-B & 0.621 & 0.970 \\
\hline
CAMP-C & 0.498 & 0.732 \\
\hline
CAMP-D & 0.092 & 0.587 \\
\hline
\end{tabular}
% \hspace{-3em}
\end{table}

\textbf{Platform level:} A large-scale online experiment with $25\%$ ADs changing to CIA bidding is conducted. The overall ROI of the ADs with CIA bidding sees an $12.1$ increase with a $4.2$ increase in cost, showing that advertiser ROI and platform revenue grow together. For all the ADs, the average increase ratio is smaller due to the performance of non-CIA bidding ADs. In sum, although our derivation of CIA lies in the perspective of the advertiser, the result shows that CIA-induced ranking has a positive influence on both advertisers and platform.
\begin{table}[htp]
\centering
\tiny
\caption{Online performance with KB as a baseline (25\% ADs switch to CIA-AD)}\label{tab:CIA-online}
\begin{tabular}{|c|c|c|c|c|}
\hline
Measures (\%) & cost & ROI & PPC & CTR \\
\hline \hline
\textbf{CIA ADs} & $\mathbf{4.2}$ & $\mathbf{12.1}$ & $\mathbf{1.8}$ & $\mathbf{1.4}$ \\ 
\hline
\textbf{All ADs} & $\mathbf{1.3}$ & $\mathbf{3.2}$ & $\mathbf{0.8}$ & $\mathbf{0.5}$ \\ 
\hline
\end{tabular}
% \hspace{-3em}
\end{table}

\section{Conclusion}\label{sec:conclusion}
In this paper, we propose a unified advertiser demands optimization framework called CIA in the e-commerce sponsored search. CIA formulates formally various demands with budget and unit conversion cost constraints. To maintain the advertiser knob on ROI, CIA constructs an impression-level bidding equation with $tk$, which bridges the advertiser's keyword-level bids and the accepted ROI. Furthermore, to optimize campaign-level demands, CIA introduces an additional cost regulator to adjust AD-level $tk$ and derives the constraint of virtual budget from the historical log. We find that under the GSP mechanism, CIA balances the benefits of three parties in sponsored search together, leading to a healthy marketing ecosystem. We illustrate the CIA optimization framework with two most representative demands as campaign-level GMV optimization and style comparison. A replay system is implemented to solve the optimization with knapsack algorithm or quadratic programming. Offline and online experimental results show that CIA provides a useful bidding strategy and is also qualified as an optimization framework for various demands.

% Considering the coupled sequential bidding processes, we will try to use attribution modeling \cite{Diemert17} to decouple the process and derive an optimal bidding strategy for future work. Furthermore, we are also trying to apply state-of-the-art reinforcement learning algorithms for this complex environment. 

% References part.
\bibliographystyle{aaai}
\bibliography{CIA-bidding-references}
\end{document}